
\input phyzzx

\def\ie{{\it i.e.}}

\def\IJMP{{\it Int.\ Jour.\ Mod.\ Phys.\ }}

\def\NP{{\it Nucl.\ Phys.\ }}
\def\PL{{\it Phys.\ Lett.\ }}
\def\PR{{\it Phys.\ Rev.\ }}

\def\PRL{{\it Phys.\ Rev.\ Lett.\ }}

\def\refmark#1{[#1]}		

\def\art#1{{\sl #1}}

\def\Exp#1{{\rm e}^{#1}}

\def\frac#1#2{{{#1}\over{#2}}}
\def\mbox#1{\mathop{#1}}

\def\half{ {1\over 2}}

\pubtype={}


\def\E{{\cal E}}
\def\A{{\cal A}}

\def\dt{{\rm d}t\,}
\def\dz{{\rm d}^2 z}
\def\DX{{\cal D}X\,}
\def\Exp#1{{\rm e}^{#1}}
\def\half{{1\over 2}}
\def\Lvec{{\vec L}}
\def\Kvec{{\vec K}}
\def\MPl{M_{\sl Planck}}
\def\pL{p^L}
\def\pR{p^R}
\def\pvec{{\vec p}}
\def\xdot{{\dot x}}
\def\xddot{{\ddot x}}

\def\zddot{{\ddot z}}
\def\zbar{\overline z}
\def\zhat{\hat z}

\def\OMIT#1{}

\def\crx{\cr\noalign{\vskip 0.2cm}}


\def\art#1{{\sl #1}}

\REF\GMa{D.J.~Gross and P.F.~Mende,
   \art{High energy behavior of string scattering amplitudes,}
   \PL {\bf 197B} (1987) 129 }

\REF\GMb{D.J.~Gross and P.F.~Mende,
   \art{String theory beyond the Planck scale,}
   \NP {\bf B303} (1988) 407}

\REF\ACV{D.~Amati, M.~Ciafaloni, and G.~Veneziano,
   \art{Superstring collisions at planckian energies,}
   \PL {\bf B197} (1987) 81;
   \art{Classical and quantum gravity effects from
   planckian energy superstring collisions,}
   \IJMP {\bf A3} (1988) 1615;
   \art{Can spacetime be probed below the string size?}
   \PL {\bf B216} (1989) 41;
   \nextline
   {G.~Veneziano,
   \art{An enlarged uncertainty principle from Gedanken
   string collisions?}
   Preprint CERN-TH.\ 5366/89}}
\REF\soldate{M.~Soldate,
   \art{Partial wave unitarity and closed string amplitudes,}
    \PL {\bf 186B} (1987) 321}
\REF\muzinich{I.~Muzinich and M.~Soldate,
   \art{High-energy unitarity of gravitation and strings,}
   \PR {\bf D37} (1988) 359 }
\REF\sundborg{B.~Sundborg,
   \art{High energy asymptotics:  the one-loop string amplitude
   and resummation,}
   \NP {\bf B306} (1988) 545 }
\REF\highsym{D.J.~Gross,
   \art{High energy symmetries of string theory,}
   \PRL {\bf 60} (1988) 1229 }
\REF\lindstrom{
   \art{The zero tension limit of the spinning string,}
    \PL {\bf B258} (1991) 331}
\REF\moore{G.~Moore,
   \art{Symmetries of the bosonic string $S$-matrix,}
   hep-th/9310026, YCTP-P19-93 }

\REF\EP{P.F.~Mende,
   \art{String theory at short distance and the principle
   of equivalence, hep-th/9210001,}
   in {\it String Quantum Gravity and Physics at the Planck
   Energy Scale, (Erice workshop, 21--28 June, 1992),}
   N.~Sanchez, ed., World Scientific 1993}

\REF\grossmanes{D.J.~Gross and J.L.~Ma\~nes,
   \art{High energy behavior of open string scattering,}
   Preprint PUPT-1116, 1989}

\REF\MO{P.F.~Mende and H.~Ooguri,
   \art{Borel summation of string theory for Planck scale
   scattering,}
   \NP {\bf B339} (1990) 641 }

\REF\greena{M.B. Green,
   \art{Space-time duality and the Dirichlet string theory,}
   \PL {\bf B266} (1991) 325}

\REF\greenb{M.B. Green,
   \art{The influence of world sheet boundaries on critical closed string
   theory, hep-th/9212085}
   \PL {\bf B302} (1993) 29}

\REF\Lia{M. Li,
   \art{Dirichlet strings,}
    hep-th/9307122, Brown-HET-915, 1993}

\REF\Lib{M. Li,
   \art{Dirichlet string theory and singular random surfaces,}
    hep-th/9309108, Brown-HET-923, 1993}

\REF\yang{Z. Yang,
   \art{Asymptotic freedom and Dirichlet string theory,}
   hep-th/9211092, UR-1288, 1992}

\REF\greenc{M.B. Green,
   \art{Temperature dependence of string theory in the presence
   of world sheet boundaries, hep-th/9201054,}
   \PL {\bf B282} (1992) 380}

\REF\BV{R. Brandenberger and C. Vafa,
   \art{Superstrings in the early universe,}
   \NP {\bf 316} (1989) 391}

\REF\TV{A.A. Tseytlin and C. Vafa,
   \art{Elements of string cosmology, hep-th/9109048,}
   \NP {\bf B372} (1992) 443}

\REF\witten{E.~Witten,
   \art{Space-time and topological orbifolds,}
   \PRL {\bf 61} (1988) 670 }

\REF\polchinski{ J.~Polchinski,
   \art{Collision of macroscopic fundamental strings,}
   \PL {\bf 209B} (1988) 252 }

\REF\khuri{R.R.~Khuri,
   \art{Veneziano amplitude for winding strings, hep-th/9303074,}
   \PR {\bf D48} (1993) 2823 }

\REF\heterotic{D.~Gross, J.~Harvey, E.~Martinec, and R.~Rohm,
   \art{Heterotic string theory~II,}
   \NP {\bf B267} (1986) 75 }

\REF\grossb{D.J.~Gross,
   \art{Superstrings and unification,}
   Preprint PUPT-1108, 1988;
   \art{Strings at superplanckian energies,}
   Preprint PUPT-1117, 1988}

\REF\lipatov{L.~N.~Lipatov,
   \art{Divergence of the perturbation theory series and
   the quasi\-classical theory,}
   {Sov.\ Phys.\ JETP} {\bf 45} (1977) 216
   ({\it Zh.\ Eksp.\ Teor.\ Fiz.} {\bf 72} (1977) 411)}

\REF\BLZ{E.~Br\'ezin, J.-C.~LeGuillou and J.~Zinn-Justin,
   \art{Perturbation theory at large order,}
   \PR {\bf D15} (1977) 1544; {\bf D15} (1977) 1558}



\date={January 1994}
\pubtype={}
\Pubnum{\vbox{ \hbox{hep-th/9401126} \hbox{Brown-HET-929} }}
\titlepage

\title{{\seventeenbf
{High Energy String Collisions in a Compact Space}
} }

\author{Paul F.~Mende
\footnote{\star}{Research supported by an SSC Fellowship, TNRLC \#FCFY9322,
and by the U.S. Department of Energy under grant DE-AC02-76-ER03130.
E-mail: {\tt mende@het.brown.edu}}}

\address{
\multiply\baselineskip by 3
\divide\baselineskip by 4
Department of Physics 			\break
Brown University			\break
Providence, Rhode Island~~02912
\smallskip}

{\abstract
\multiply\baselineskip by 3
\divide\baselineskip by 4
When high energy strings scatter at fixed angle,
their amplitudes characteristically fall off
exponentially with energy, ${\cal A} \sim \exp(-s \times const.)$.
We show that in a compact space this suppression disappears for
certain kinematic configurations.
Amplitudes are power-law behaved and therefore greatly enhanced.
In spacetime this corresponds to fixed-angle scattering, with fixed
transfer in the compact dimensions.  On the worldsheet this process
is described by a stationary configuration of effective charges and
vortices with vanishing total energy.
It is worldsheet duality---and not spacetime duality---that plays
a role.
\smallskip}

\vfil
\eject

{\bf\chapter{\bf Introduction}}

Strings differ from particles in two crucial respects:
at short distance they have internal structure, and at long distance they
are sensitive to the global structure of spacetime.
The first property is fundamental to a meaningful
theory of quantum gravity and shows up in high energy processes.
The second plays a role in compactifications,
symmetry breaking, and thermodynamics.
The first derives from the tower of massive string modes;
the second derives from string zero modes.

In this paper we study scattering amplitudes for string winding
states at high energy, and show that
new behavior---absent for unwound strings---can arise
for certain configurations.
Compactifying dimensions on an internal manifold of
size of the order of the Planck length or larger has many well-known
advantages for low-energy phenomenology, and is obviously
important if we hope to use critical string theory to describe
our four-dimensional world.
The question we address here is how the existence of compactified
dimensions affects the interactions of the string at its natural
scale.

The high-energy limit of string theory is not only the most interesting
place to look for ``stringy'' effects likely to be characteristic
of quantum gravity, but also a limit which can be analyzed in great detail
\refmark{\GMa, \GMb, \ACV, \soldate, \muzinich, \sundborg,
\highsym, \lindstrom, \moore}.
\foot{Small low energy effects signalling internal structure
and affecting the equivalence principle were discussed in \refmark{\EP}.}
The amplitudes for closed
and open\refmark{\grossmanes} string scattering can be computed to
all orders in perturbations theory, and the full perturbation theory
can be resummed in closed form
using Borel techniques\refmark{\MO}.
Moreover, high energies yield the semi-classical limit of string theory,
and many interactions can be understood in terms of a ``master trajectory''
which solves the classical equations of motion\refmark{\MO}.

More recently, there has been greatly renewed interest in formulating
QCD as a string theory.   One of the obstacles to this
program has always been that while string theory shares the
Regge behavior observed of hadrons in fixed momentum transfer
scattering, it departs dramatically in fixed angle scattering.
Indeed, the very softness of the string that makes it such
a successful quantum extension of general relativity makes
it unable to describe parton-like behavior:
string amplitudes fall exponentially with energy instead of
as a power law.
Had QCD not been developed this fact would have
been enough to eventually kill off strings as the theory of strong
interactions.

Green has proposed a clever generalization of the ordinary bosonic
string\refmark{\greena,\greenb} in which boundaries are included
on the string world sheets.  The Dirichlet conditions imposed
on these boundaries effectively map a finite part of the world
sheet onto a single spacetime point, thereby creating
pointlike interactions and a power law amplitude at high energies,
while the long-distance behavior is unaffected.
This proposal, still under
investigation\refmark{\Lia, \Lib, \yang, \greenc},
has made clear the importance of understanding the full
range of high energy behaviors that could arise from string theory.

The power law amplitude described below represent a totally
different way of obtaining non-exponential behavior, though
not one likely to have application to QCD ---
the origin has no obvious connection with short distance
spacetime physics.  Rather, as it comes about in
a compactified space, it may be of relevance of string
thermodynamics and cosmology (e.g., in extending the cosmological
calculations
of \refmark{\BV,\TV} to include the detailed cross
sections to account for unusually sharp energy and angle
dependence).

It has been thought for some time that string theory might contain a more
symmetric structure, visible at high energies, and which is realized
in a broken-symmetry phase in the conventional
formulation  (as advocated, e.g., in \refmark{\witten, \highsym}).
We do not offer any decisive evidence for or against this interesting
conjecture, but hope by the present study to become more familiar
with the structure of string theory in this decidedly unfamiliar energy
regime.


Recall that for the scattering of tachyons in the bosonic string, the
$G$-loop amplitude is given by
$$
   \A_G(p_i) = \int {\cal D}g {\cal D}X  \Exp{-I[X,g] +i\int P\cdot X}
   \eqn\pathsum
$$
where $X^\mu(\xi)$ is the spacetime trajectory of the string;
the string action is the invariant area of the worldsheet,
$$
   I[X,g] = -{1 \over 2\pi} \int {{\rm d}^2\xi}
   \sqrt{g} g^{\alpha\beta} \partial_{\alpha}X^{\mu}\partial_{\beta}X_{\mu} ;
   \eqn\?
$$
$g_{\alpha\beta} $ is the metric on the punctured worldsheet of genus~$G$;
and $P$ is a source for the particles,
$ P^\mu (\xi) = \sum p^\mu_i \delta^{(2)}(\xi,\xi_i). $

When all of the momenta get large 
the path integral is dominated by a classical trajectory and
can be studied semiclassically\refmark{\GMa, \GMb}.
This is easy to see:
define rescaled string coordinates and momenta by
$$
   {\tilde X} = {X\over \sqrt{s} },
   \qquad
   {\tilde p}_i = {p_i \over \sqrt{s}}
   .\eqn\Xscaled
$$
Then
$$
   \A_G(s;{\tilde p}_i) = \int {\cal D}g {\cal D}{\tilde X}  \,
   \exp\left\{ s\left(-I[{\tilde X},g]
   +i\int {\tilde P}\cdot {\tilde X}\right) \right\}.
   \eqn\?
$$
Thus the $s\to\infty$, or equivalently
$\MPl \to 0$, limit corresponds to
the semiclassical limit of string theory\refmark{\highsym},
with the dominant contribution coming from the trajectory
$$
   {\tilde X}^\mu(\xi)
   = \sum_i {\tilde p}^\mu_i \, G(\xi,\xi_i)
   \eqn\Xclass
$$
where $G(\xi,\xi')$ is the Green function for the action $I[{\tilde X},g]$.

Evaluating the action at the stationary point gives the dominant behavior,
$$
   \A_G \sim \exp\left(-{s \over G+1} \right)
   ,\eqn\?
$$
for the amplitude at genus $G$.
The exponent is the effective electrostatic energy
of a collection of ``charges'' $p_i$ at the points $\xi_i$ on the worldsheet,
$$
   \E = -{1\over 2}\sum p_i\cdot p_j G(\xi_i,\xi_j)
   .\eqn\?
$$


There are two reasons one might naively
believe that compact dimensions do not
affect the basic picture of high-energy interactions.
First, if we scatter any fast-moving states
with energies $E$ much greater than
$\MPl$, we expect that they will probe short wavelengths much less
than the compactification radius, typically $\lsim 1/\MPl $.
That is, at high energies the compact dimensions look big, and the
strings are not sensitive to the compactification.

One can see this in more detail from the following observation.
If the background is a product of flat non-compact spacetime with an
{\it arbitrary\/}
conformal field theory of appropriate central charge, the vertex operators
are of the form
$$
   \int\dz\,e^{ip\cdot X} \, {\cal P}(\partial^m X, {\overline\partial}^n X)
   \, \Phi(z,\zbar),
   \eqn\?
$$
where $\Phi$ is an operator in the conformal field theory and $\cal P$ is
a polynomial in the field derivatives.
In the high-energy limit, the momentum dependence arising from $\cal P$
insertions does not shift the stationary point, and its correlators
as well as those
of $\Phi$, which are momentum-independent, can simply be evaluated
at the stationary point, which depends only on the universal factor
$e^{ip\cdot X}$.
Thus only the non-universal prefactors are modified.

The second reason is that
it might seem that taking a large winding number
for the scattering states
is equivalent to letting the radius of the compact dimensions go to
infinity.  Then not only is the higher dimensional space recovered,
but the symmetry group as well.  This means that Lorentz generators
could mix internal momenta with the time direction, and the
internal kinematics then becomes trivial since one can boost to
a center of mass frame.
We will see how these expectations fall short of the mark.

In the next section we show that for certain winding configurations
the fixed-angle scattering amplitudes are exponentially enhanced.
This follows from simple properties of the world sheet physics.
In the following section, we describe the spacetime kinematics in
detail.  Finally, we offer a discussion of the results and some
speculations on possible interpretation.

{\bf \chapter{\bf Gone with the wind}}

Consider now a bosonic string moving in a space
with some dimensions compactified  on a torus.
The vertex operator for the creation of a moving, wound string is
$$
   V(\pL, \pR) = \int \dz\,
   \exp\left\{ {i\pL\cdot X(z) + i\pR\cdot X(\zbar)} \right\}
   ,\eqn\windingvertex
$$
where the left and right momenta are defined in terms of the momentum
and winding vectors,
$\pL_\mu={1\over 2}(p_\mu + L_\mu)$, $\pR_\mu={1\over 2}(p_\mu-L_\mu)$.
Therefore the amplitude, at tree level, for the scattering of $N$
strings is of the form of the integral of a holomorphic times an
antiholomorphic function,
$$
   \eqalignno{
   \A &= \int\prod_i\dz_i\,
   \prod_{i<j} (z_i-z_j)^{\pL_i\pL_j}
   \, \prod_{i<j} (\zbar_i-\zbar_j)^{\pR_i\pR_j}
   & \eqname\amp
   \cr
   &= \int\prod_i\dz_i\,
   \prod_{i<j} |z_i-z_j|^{p_ip_j/2 + L_iL_j/2}\,  \prod_{i<j}
   \left({z_i-z_j\over \zbar_i-\zbar_j}\right)^{p_iL_j+p_jL_i}
   .& \eqname\EM
   \cr}
$$

The amplitude has the form
$$
   \A = \int\prod\dz_i\,\exp\left\{\, -\E_L(z)-\E_R(\zbar) \,\right\}
   .\eqn\?
$$
Let us recall that one can regard the effective action
$\E_L + \E_R$ as the electromagnetic
energy of a collection of charges and vortices on
the worldsheet, where at point $z_i$ there is an object with
electric charge~$p_i$ and magnetic charge~$L_i$, and
$$
   \eqalign{
   \E_L(z) &= -\sum_{i<j} \pL_i\cdot\pL_j\log(z_i-z_j)
   ,\crx
   \E_R(\zbar) &= -\sum_{i<j} \pR_i\cdot\pR_j\log(\zbar_i-\zbar_j)
   .\cr
   }\eqn\ELERdef
$$
(The fact that $p_i$ and $L_i$ are vectors rather than real numbers,
and furthermore can have non-positive inner product, will haunt us
shortly.  The picture is nonetheless helpful.)

Electromagnetic duality on the two-dimensional worldsheet results in
the symmetry of Eq.~$\EM$ under the formal interchange of windings and
momenta, $p_i \leftrightarrow L_i$.
Observe that the energy of a set of pure magnetic vortices ($p_i=0$)
has the same form as the energy of a set of pure
electric charges ($L_i=0$).
This symmetry is only formal,
however: since time is not compact
the mass-shell condition
(required for conformal invariance) breaks it.
In other words, this duality is not useful for scattering
amplitudes because it gives relations among unphysical processes.

As we pass to the high-energy limit,
the introduction of winding states dramatically changes the physics.
When {\it all\/} of the momentum invariants,
$\pL_i\cdot \pL_j$ and~$\pR_i\cdot \pR_j$,
get large simultaneously,
the integral is dominated by a stationary point, given by
$$
   \eqalign{
    {\partial\E_L(z)\over\partial z_i} &=0  ,
   \qquad
    {\partial\E_R(\zbar)\over\partial \zbar_i} =0 .
   \cr }
   \eqn\sadpt
$$

This system of equations is in general
overdetermined.
Witten\refmark{\witten}
observed that in the general case
a stationary point exists only if one takes the coordinates~$z_i$
and~$\zbar_i$ to be independent complex variables,
rather than complex conjugates,
leading to a description of the amplitude in terms of independent
moduli for the left and right moving degrees of freedom of the string.

Consider the physics of several special cases.
If the strings are all unwound,
$L_i=0$, then $\E_L=\E_R$ and Eqs.~$\amp$ and $\EM$ reduce
to the familiar Koba-Nielsen form: the
integrand is the absolute value of an analytic function.
Eqs.~$\sadpt$ now have a solution since they are identical.

Observe that this also occurs if the momenta and windings
are chosen such that
$p_i\cdot L_j=0$ for all~$i$ and~$j$.
The second factor of~$\EM$ drops out and again
$\pL_i\cdot\pL_j = \pR_i\cdot\pR_j$.
This case was considered
by Polchinski, who used $\amp$ to compute the
total cross-section for macroscopic fundamental strings\refmark{\polchinski}.
(Winding string amplitudes were also studied by Khuri\refmark{\khuri}.)

Another possibility is to take self-dual windings:
Then~$\pR_i=0$, $\E_R=0$, and the integrand is holomorphic, involving the
integrand that appears for the open string.
Witten identified the string trajectories of
this case with the instantons of the two-dimensional
topological sigma model\refmark{\witten}.
Unfortunately this intriguing case cannot be realized without
compactifying time or violating the mass-shell condition.

Consider now what happens if one sets
$$
   \pL_i\cdot\pL_j =-\pR_i\cdot\pR_j,
   \eqn\alie
$$
which makes the first factor of~$\EM$ drop out.
The integrand is pure phase,
a holomorphic function divided by its conjugate.

For such strings two remarkable facts emerge:
\item{\it (1)\/}~there is a real solution to the saddle point condition
Eq.~$\sadpt$,
and \item{\it (2)\/}~at the stationary point,
{\it the string action is zero!\/}

Indeed, using~$\alie$ in~$\ELERdef$ we have
$$
   \E_L = -\E_R \equiv \E
   ,\eqn\?
$$ in which case the effective action
$\E(z) - \E(\zbar)$ looks purely magnetic in form.
\foot{
More generally, Eq.~$\sadpt$ has solutions if the left and right
energies are
proportional, $\E_R = \alpha \E_L$.  For momentum states and for the strings
of ref.~\refmark{\polchinski}, $\alpha=1$; for the self-dual case of
ref.~\refmark{\witten}, $\alpha=0$; and for the magnetic case here,
$\alpha=-1$.
}

The amplitudes can be read off from comparison with pure
tachyon scattering amplitudes, since only the sign of~$\E_R$ differs.
The equations~$\sadpt$ determining the stationary point are identical,
as are the (real) functions~$\E_L(z)$
and~$\E_R(\zbar)$ evaluated at the stationary point.
But instead of adding in the exponent, they cancel:
$\E_L+\E_R=0,$ and there
is no exponential dependence on the energy at all:
$$
   \A \sim \Exp{-\E_L - \E_R} \sim {\cal O}(1)
   .\eqn\?
$$

The amplitude is therefore dominated by the prefactors and the
fluctuations around the saddle point, and behaves
as a power law function of the momenta only.

{\bf \chapter{\bf Kinematic considerations}}

To be concrete, consider four-point scattering
of bosonic string states given by~$\windingvertex$.
We take the 26 dimensions of the string background to be a product space
of Minkowski space and a toroidal internal space.
\foot{
The respective
dimensions are unimportant, provided that the internal space
has dimension at least equal to six.
With fewer dimensions,
there are additional constraints on the number of independent
internal kinematic invariants that can be formed.
}

First we establish kinematic notation:
Let the momentum of a string  be
$p = (E, \pvec, \Kvec)$, where $E$ is the energy, $\pvec$ is
the momentum in the non-compact dimensions, and $\Kvec$ is the internal
momentum vector.  The winding vector of the string is $L=(0,0,2\Lvec)$.
(The unconventional factor of 2 is introduced
here to keep the formulas symmetric.)

Since there are no oscillators excited in these states, the mass shell
condition for the state reads
$$
   (\pL)^2 = (\pR)^2 = 2.
   \eqn\massshell
$$
{}From $(\pL)^2=(\pR)^2$, we learn that $\Kvec\cdot\Lvec = 0$.
{}From $(\pL)^2=2$, we get the equation for the mass:
$p^2 + L^2 = 8$.
To an observer in the uncompactified space the energy is given by
$$
   ({\rm mass})^2 = \Kvec^2 + \Lvec^2 -8,
   \eqn\?
$$
which grows as the components of $\Kvec, \Lvec$ increase.

Define left and right generalizations of the usual Mandelstam
variables in the obvious way: $s_L = -(\pL_1 + \pL_2)^2$, \etc.
They satisfy 
$$
   s_L+t_L+u_L = - 8 = s_R + t_R + u_R.
   \eqn\?
$$
Then let $S \equiv s_R-s_L = p_1\cdot L_2 + p_2\cdot L_1$,
and similarly for $T$ and $U$, which satisfy
$$
   S+T+U=0.
   \eqn\?
$$
Finally, let
$\lambda_{ij}$ be the angle between~$\Lvec_i$ and~$\Lvec_j$,
$\kappa_{ij}$ be the angle between~$\Kvec_i$ and~$\Kvec_j$, and
$\pi_{ij}$ be the angle between~$\pvec_i$ and~$\pvec_j$.

Now it is simple algebra to show that
Eq.~$\alie$ is actually impossible because the winding vectors have purely
Euclidean signature.
It is equivalent to
$$
   p_i\cdot p_j \, + \, L_i\cdot L_j = 0,
   \eqn\anotherlie
$$
which can only be satisfied if some of the lengths $|\pvec|<0$.

Nevertheless the saddle point problem is unaffected if we
can satisfy Eqs.~$\alie$ and~$\anotherlie$
approximately,
$$
   \pL_i\cdot\pL_j \, + \,  \pR_i\cdot\pR_j \, \approx \,0
   ,\eqn\?
$$
so that in the analog electromagnetic
problem the energy is held finite as the charges
become infinite.
To be precise, let
$$
   \pL_i\cdot\pL_j \, + \,  \pR_i\cdot\pR_j \,=\, c_{ij},
   \eqn\constraint
$$
where the magnitude of each term on the left hand side goes to infinity,
and the $c_{ij}$ are constants satisfying
({\it i\/}) $c_{12}<0$;
({\it ii\/}) $c_{13}, c_{14} > -4$;
and ({\it iii\/}) $c_{12}+c_{13}+c_{14}=-4$.

This can be done in various ways.  For example,
we let the magnitudes of
the internal vectors~$|\Kvec_i|$, $|\Lvec_i|\to\infty$,
while the angles are scaled as
$\lambda_{12}^2 \sim -2 c_{12}/(|\Lvec_1||\Lvec_2|)$,
$\kappa_{12}^2 \sim -2 c_{12}/(|\Kvec_1||\Kvec_2|)$.
The $t$-channel
angles scale as
$\lambda_{13} \sim \pi - \left((2c_{13}+8)/(|\Lvec_1||\Lvec_3|)\right)^{1/2}$,
$\kappa_{13} \sim \pi - \left( (2c_{13}+8)/(|\Kvec_1||\Kvec_3|)\right)^{1/2}$.

Using this parameterization of the limit (and~$SL(2,C)$
invariance to fix three variables as usual) the amplitude is
$$
   \eqalign{
   \A &= \int\dz\, |z|^{c_{13}} |1-z|^{c_{14}}
   \left( {z\over\zbar} \right)^{T}
   \left( {1-z\over 1-\zbar} \right)^{U}
   \crx
   &= \int\dz\, |z|^{c_{13}} |1-z|^{c_{14}}
   \exp -S\left\{ {T\over S}\log{z\over\zbar}
   + {U\over S}\log{ 1-z\over 1-\zbar} \right\}
   .\cr}
   \eqn\fourpoint
$$

In the high-energy limit
(that is, $S$, $T$, $U\to\infty$, while $T/U$, $c_{ij}$,
\etc, are fixed) the integral is dominated by the stationary point at
$\zhat = -T/S$, yielding
\foot{At tree level,
a four-point function of the form~$\amp$ can also be evaluated using a
formula of Ref.~\refmark{\heterotic, appendix A}
}
$$
   \A = S^{-1} {\zhat  (1-\zhat)}
   |\zhat|^{c_{13}} |1-\zhat|^{c_{14}}
   ,\eqn\largeS
$$
up to corrections of
$ {\cal O}(1/S^2)$,
or in crossing-symmetric form
$$
   \A = (STU)^{-5/3} \left|S^{c_{12}}T^{c_{13}} U^{c_{14}}\right|
   .\eqn\largeStoo
$$

{\bf\chapter{\bf Discussion}}

The result $\largeS$ should be compared with the behavior of high-energy
scattering of unwound strings, $\A\sim\exp(-s)$ for fixed angle, and
$\A\sim s^{2+t/4}$ for fixed $t$.   The result superficially
resembles the Regge behavior of the latter.
But like the former, the dominant
contribution comes from regions in the middle of the moduli space
(\ie, from non-degenerate Feynman graphs).
For these scattering processes, the amplitude is greatly
enhanced relative to the characteristic exponential string falloff.

This result may be extended to higher genus as well, along the lines
of \refmark{\GMa, \GMb}.
The stationary condition will apply as well
on Riemann surfaces of $N=G+1$ sheets, of the form
$$
   y^N=\prod_{i=i}^4(z-z_i)^{L_i}
   \eqn\ycurve
$$
That is, precisely the same Riemann surfaces are saddle points for
both the winding state and momentum state amplitudes.
\foot{
We do not have a simple
argument to show that these stationary points are dominant,
as for the case of unwound strings, but we
assume this to be the case.
}
Now observe that the operators located at the branch points of the
Riemann surface behave as vortex operators, rather than electric charges,
but by electromagnetic duality, the electrostatic and magnetostatic
configurations are simply related.

The rest of the calculations
of \refmark{\GMa, \GMb} may be carried
out, with the principal difference arising in the phase of the determinant
of fluctuations about the stationary point, which now changes sign,
so that all terms in the perturbative series appear with the same sign.
It was precisely this sign which made the high-energy amplitude
Borel summable\refmark{\MO}, and this property is now lost.

To further explore the physics of strings in Kaluza-Klein spacetimes,
it would be interesting to compute, for example, branching ratios for
the production of high energy states into wound versus unwound states.
This cannot be done for the case of the four-point amplitudes:  only
trivial winding/antiwinding pairs can be produced, so it is necessary to
go to higher point functions.  The kinematics for two initial unwound
strings requires that $s_L=s_R$.

In considering higher point functions, other possibilities emerge as well.
(Such amplitudes have recently been explored by Moore\refmark{\moore}.)
For four points, the saddle point equations are linear and have a single
solution.  For $M$ points, they are degree $M-3$, with as many solutions.
It might be possible to consider cases where different solutions to the left
and right equations are combined.

The spacetime trajectory corresponding to the scattering is given by
$$
   X^\mu(z,\zbar) = {i\over 2} \sum_{i=1}^4 \left( p_i^\mu \log |z-a_i|^2
   + L_i^\mu \log\left({z-a_i\over \zbar - {\overline a}_i}\right)\right)
   .\eqn\traj
$$

How are we to understand this behavior?  Gross\refmark{\grossb} has suggested
that the exponential behavior of the fixed-order amplitudes,
$$
   \A(s)\sim \Exp{-\alpha'({\rm const.})s}
   ,\eqn\?
$$
indicates a suppression typical of a classically forbidden process,
$$
   \A\sim \Exp{-I/\hbar}
   .\eqn\?
$$

If this is the case, then the scattering of winding states corresponds to
{\it classically allowed\/} processes in spacetime.

How might string trajectories shed light on the conjectured
topological phase of string theory?
An analogy from quantum mechanics may be suggestive.
Suppose we wish to compute the ground state energy of the
anharmonic oscillator,
$$
   V(x) = {1\over 2}x^2 + gx^4
   ,\eqn\?
$$
from the Euclidean functional integral
$$
   Z(T,g) = \int\DX\exp\left\{ -\int_0^T\dt
   \left(\half\xdot^2 + \half x^2 + gx^4\right) \right\}
   .\eqn\?
$$
As $T\to\infty$, $E_0 \sim -(\log Z)/T$.
One can expand perturbatively in powers of the coupling $g$ and
following \refmark{\lipatov, {\BLZ}}
exponentiate the interaction vertices to
write
$$
   \eqalign{
   Z(T,g) &= \sum_{G=0}^\infty { (-g)^G \over G! } Z_G(T)
   ,\cr
   Z_G(T) &= \int\DX\exp\left\{ -I_0 + G\log\int_0^T\dt x^4 \right\}
   .\cr}
   \eqn\Expansion
$$
At large order, the path integral $Z_G$ is dominated by a saddle point
trajectory where the effective action is stationary:
$$
   \xddot = x - 4x^3{G\over J[x]}
   \eqn\eqmotion
$$
where $J[x]\equiv \int_0^T \dt x^4$.  By rescaling,
$$
   x = z\left( {J[x]\over G} \right)^{1/2} = z\left({G\over J[z]}\right)^{1/2}
   \eqn\?
$$
the equation of motion for the trajectory {\it becomes independent of $G$}:
$$
   \zddot = z - 4z^3
   .\eqn\Saddlept
$$
The resulting perturbative series for $Z(T,g)$ is divergent,
but of alternating signs and hence Borel summable.
(The origin of the divergence is the illegal change of order of summation
and path integration in Eq.~$\Expansion$.  For fixed $g$, large values
of $x(t)$ always dominate, so one cannot simply integrate term by term.)
Therefore we find the following situation, analogous to string theory
(cf. \refmark{\GMb, \MO}:
\medskip
\point
The amplitude is expressed as a perturbative sum of path integrals.
\point
Each integral is dominated by a saddle point trajectory.
\point
These trajectories {\it are the same at each order of perturbation theory\/},
up to a scale.
\point
The resulting series is divergent, but Borel summable!
\medskip

Now in particle quantum mechanics we have the Lagrangian and a
well-defined non-perturbative formulation of the amplitude, so we can
see what features of the full theory are
suggested by the perturbative analysis.

In the full theory the potential is
$V(x) = {1\over 2}x^2 + gx^4$,
and the Euclidean equation of motion  is
$$
   \xddot = x+4gx^3
   .\eqn\?
$$
For $g<0$, this is the saddle point equation $\Saddlept$.
Therefore the saddle point trajectories are the instantons of the
opposite phase of theory.
Perhaps then, as hoped in tree-level comparison
with topological sigma models of ref.~\refmark{\witten}, similar relations
hold in string theory and might be elucidated by further
investigation of spacetime string trajectories.

 \refout
\end